\documentclass[journal=jacsat,manuscript=article]{achemso}
\usepackage{setspace}
\usepackage[utf8]{inputenc}
\usepackage[T1]{fontenc}
\usepackage{bm}
\usepackage{color}

\usepackage{ulem}
\usepackage{xcolor} % more colors
\usepackage{hyperref} % for jumping to references

\usepackage{siunitx}
    \DeclareSIUnit\angstrom{\text {Å}}
    \DeclareSIUnit\bar{bar}
\author{Rahul Sharma}
\altaffiliation{These authors contributed equally to this work.}
\affiliation{Univ. Grenoble Alpes, CEA, CNRS, IRIG-SPINTEC, 
38000 Grenoble, France}

\author{Sylvain Massabeau}
\altaffiliation{These authors contributed equally to this work.}
\affiliation{Laboratoire Albert Fert, CNRS, Thales,
Universit\'e Paris-Saclay, F-91767 Palaiseau, France}

\author{Armando Pezo}
\altaffiliation{These authors contributed equally to this work.}
\affiliation{Laboratoire Albert Fert, CNRS, Thales,
Universit\'e Paris-Saclay, F-91767 Palaiseau, France}

\author{Ekta Yadav}
\affiliation{Laboratoire Albert Fert, CNRS, Thales,
Universit\'e Paris-Saclay, F-91767 Palaiseau, France}

\author{Viliam Vretenár}
\affiliation{Centre for Nanodiagnostics of Materials, Faculty of Materials 
Science and Technology, Slovak University of Technology, Vazovova 5, 
Bratislava, 812 43, Slovakia}

\author{Ravi K. Biroju}
\affiliation{Centre for Nanodiagnostics of Materials, Faculty of Materials 
Science and Technology, Slovak University of Technology, Vazovova 5, 
Bratislava, 812 43, Slovakia}
\alsoaffiliation{Nanoelectronics \& VLSI Design and Department of Physics, 
School of Advanced Sciences, Vellore Institute of Technology, 
Chennai 600127, Tamil Nadu, India}

\author{Fatima Ibrahim}
\affiliation{Univ. Grenoble Alpes, CEA, CNRS, IRIG-SPINTEC, 
38000 Grenoble, France}

\author{Sukhdeep Dhillon}
\affiliation{Laboratoire de Physique de l’Ecole normale sup\'{e}rieure, ENS, Universit\'{e} PSL, CNRS, Sorbonne Universit\'{e}, Universit\'{e} de Paris, 24 rue Lhomond, 75005 Paris, France}

\author{Alain Marty}
\affiliation{Univ. Grenoble Alpes, CEA, CNRS, IRIG-SPINTEC, 
38000 Grenoble, France}

\author{Isabelle Gomes de Moraes}
\affiliation{Univ. Grenoble Alpes, CEA, CNRS, IRIG-SPINTEC, 
38000 Grenoble, France}

\author{Adrien Michon}
\affiliation{Universit\'e C\^ote d'Azur, CNRS, CRHEA, rue Bernard 
Gr\'egory, Valbonne, France}

\author{Jing Li}
\affiliation{Univ. Grenoble Alpes, CEA-LETI, 38000 Grenoble, France}

\author{Mairbek Chshiev}
\affiliation{Univ. Grenoble Alpes, CEA, CNRS, IRIG-SPINTEC, 
38000 Grenoble, France}

\author{Henri Jaffr\`es}
\affiliation{Laboratoire Albert Fert, CNRS, Thales,
Universit\'e Paris-Saclay, F-91767 Palaiseau, France}

\author{Jean-Marie George}
\affiliation{Laboratoire Albert Fert, CNRS, Thales,
Universit\'e Paris-Saclay, F-91767 Palaiseau, France}

\author{Matthieu Jamet}
\affiliation{Univ. Grenoble Alpes, CEA, CNRS, IRIG-SPINTEC, 
38000 Grenoble, France}

\email{matthieu.jamet@cea.fr}

\title{Thickness-Dependent Spintronic Terahertz Emission in MBE-Grown PtTe\textsubscript{2}: From Semiconductor to Type-II Dirac Semimetal}

\begin{document}
\vspace{1cm}

% -------------------------------------------------------
% ABSTRACT 
% -------------------------------------------------------
\newpage
\begin{abstract}
Spintronic terahertz (THz) emitters have established themselves as among the most practical broadband THz sources available, yet their performance remains fundamentally limited by the spin Hall conductivity of the nonmagnetic conversion layer --- a quantity that is fixed once the material is chosen. Here, we demonstrate that in PtTe\textsubscript{2}, a type-II Dirac semimetal within the transition metal dichalcogenide family, this limitation can be circumvented by exploiting the dramatic thickness-driven electronic phase evolution of the material itself. Using molecular beam epitaxy to grow PtTe\textsubscript{2} films with single-monolayer precision from 1 to 20\,ML, we show that the spintronic THz emission tracks the underlying electronic phase diagram directly: it is absent in the single-layer semiconducting phase, turns on sharply at the semimetal transition near 2\,ML, and reaches a peak amplitude six times that of an equivalent Pt reference at 10\,ML, before declining at larger thicknesses due to THz reabsorption in the increasingly metallic film. This non-monotonic behavior is inconsistent with a bulk inverse spin Hall mechanism and instead reflects a multi-channel spin-to-charge conversion process in which spin-momentum-locked topological surface states and a thickness-dependent interfacial Rashba splitting both contribute and strengthen as the type-II Dirac band structure develops. First-principles calculations of the interfacial spin accumulation reproduce the experimental trend quantitatively, confirming this physical picture. These findings introduce thickness engineering of van der Waals semimetals as a new and accessible route to optimizing spintronic THz emitters and spin-orbit torques in magnetic memories (SOT-MRAMs), with direct implications for the broader class of dimensionally tunable topological materials.
\end{abstract}

% -------------------------------------------------------
% KEYWORDS
% -------------------------------------------------------
\noindent\textbf{Keywords:} type-II Dirac semimetal, spintronic terahertz emitter, spin-to-charge conversion, topological surface states, van der Waals materials

% -------------------------------------------------------
% INTRODUCTION
% -------------------------------------------------------
\newpage
\section*{Introduction}

Spintronic THz emitters (STEs), thin-film heterostructures pairing 
a ferromagnetic (FM) spin source with a nonmagnetic (NM) spin-to-charge 
conversion layer, have established themselves as among the most practical 
and scalable broadband THz sources available today~\cite{Kampfrath2013}. 
The intensity of the emitted THz pulse is directly proportional to the 
spin Hall conductivity (SHC) of the NM layer, which is a key figure of 
merit for STE performance~\cite{Liu2011}. Pt has long dominated this 
role because of its combination of high electrical conductivity and a 
spin Hall angle of $\sim$0.06-0.08, yielding a SHC of 
$\sim$3.5-3.8\,$\times$\,$10^{5}$\,($\hbar/2e$)($\Omega\cdot$m)$^{-1}$
~\cite{Zhu2019, Ganguly2014, Liu2011}. The requirement of scalable 
heterostructures with tunable functionalities has driven the exploration 
of alternative NM materials. Topological insulators such as 
Bi$_{2}$Se$_{3}$ offer large spin Hall angles and Rasbha-Edelstein effect through spin-momentum-locked 
topological surface states, yet their bulk resistivity is orders of 
magnitude higher than Pt, thereby limiting the SHC to 
$\sim$1-2\,$\times$\,$10^{4}$\,($\hbar/2e$)($\Omega\cdot$m)$^{-1}$ 
and yielding only modest THz emission in 
practice~\cite{Wang2018, Tong2021}. The central challenge is therefore 
not simply to find a large spin Hall angle, but to find it in a material 
that is also highly conductive - a combination that has proved elusive 
across most known quantum materials.

Transition metal dichalcogenides (TMDs) have attracted growing interest as spintronic NM layers due to their exceptional tunability in electrical conductivity, spin--orbit coupling (SOC) strength, and topological band structure.\cite{ Xu2014,Yang2022} Their layered van der Waals nature allows precise thickness control and clean interface engineering, which are critical for optimizing spin injection and spin-to-charge conversion (SCC).\cite{Abdukayumov2024,Yang2022} Yet when it comes to spin-to-charge conversion for THz emission, most metallic TMDs face a practical ceiling. WTe\textsubscript{2}, NbSe\textsubscript{2}, and MoTe\textsubscript{2} can exhibit respectable spin Hall angles, but their electrical conductivities are several orders of magnitude below those of conventional heavy metals, and since the spin Hall conductivity — the quantity that directly sets THz emission amplitude — is the product of the two, the net result has been underwhelming.\cite{Xu2020,Stiehl2019}

PtTe$_{2}$ is a notable outlier here. Classified as a type-II Dirac semimetal, it hosts strongly tilted Dirac cones that violate Lorentz invariance and give rise to topological surface states intersecting the Fermi level between the $\Gamma$ and M points~\cite{Yan2017, Politano2018}. More practically, it carries the highest room-temperature electrical conductivity of any known metallic TMD, around $3.3\times 10^{6}$\,S\,m$^{-1}$, which alone makes it interesting for spintronics~\cite{Hao2018, Fu2018, Yadav2024}. This high conductivity coexists with multiple independent spin-to-charge conversion channels: the bulk inverse spin Hall effect, whose efficiency is comparable to Pt; an  inverse Rashba- Edelstein effect contribution from the spin-momentum-locked surface states; and a local Rashba channel originating from the internal dipole field between the Pt and Te sublattices. The combined effect is a spin Hall conductivity of $0.2$--$2\times 10^{5}$\,($\hbar/2e$)($\Omega\cdot$m)$^{-1}$, the largest reported among TMDs and squarely in the range of topological insulator~\cite{Yadav2024, Xu2020}. This translates directly into THz emission: PtTe$_{2}$/Co bilayers have been shown to surpass optimized Pt/Co emitters by $\sim$15\% in peak amplitude, and to nearly double the output of a same-thickness Pt reference~\cite{Yadav2024}.

A dimension of PtTe\textsubscript{2} physics that has thus far received little attention in the context of THz emission is its thickness-driven electronic phase evolution. Unlike most TMDs, which undergo only modest changes in band structure upon thinning, PtTe\textsubscript{2} displays a strong interlayer coupling mediated by Te~5$p$ orbital overlap across the van der Waals gap.\cite{Lin2020} Angle-resolved photoemission spectroscopy (ARPES) measurements combined with first-principles calculations have established a clear electronic phase diagram as a function of thickness: bulk and multilayer PtTe\textsubscript{2} ($\geq 5$ monolayers, ML) is semimetallic, with bands that cross the Fermi level and a negative electronic gap that opens and deepens as more layers are added; at two ML, the material is a semimetal; at the single-ML limit, a Lifshitz transition converts the material into a semiconductor with a large positive indirect gap of $+$0.79~eV.\cite{Lin2020,Yan2017} This dimensionality-mediated semimetal-to-semiconductor transition, driven by the suppression of interlayer Te~5$p$ coupling, which is among the most drastic electronic phase changes observed in any van der Waals material family. Furthermore, as thickness increases beyond the few-layer semimetal regime, PtTe\textsubscript{2} films develop progressively more bulk-like character and ultimately recover the full three-dimensional type-II Dirac semimetallic band structure with its associated topological surface states and large SHC. The interplay between this rich thickness-dependent phase evolution — semiconductor $\to$ thin-film semimetal $\to$ type-II Dirac semimetal and the resulting spintronic THz emission properties remains a largely unexplored domain.

In this work, we present a systematic study of spintronic THz emission in MBE-grown PtTe\textsubscript{2} thin films spanning the complete thickness range from 1 to 20~ML. The layer-by-layer precision of MBE growth allows us to traverse the full electronic phase diagram of PtTe\textsubscript{2} in a single material system and directly correlate the THz emission response with the underlying electronic transitions. We experimentally show that THz emission is negligible in the single-ML semiconducting phase, turns on sharply at the semimetal transition near 2~ML, and increases continuously with thickness as the Dirac surface states and bulk spin--orbit coupling of the type-II semimetal phase are progressively restored. First-principles calculations of the thickness-dependent spin Hall conductivity reproduce this trend quantitatively and identify the microscopic origin of each enhancement stage. The correspondence between the well-established electronic phase diagram of PtTe\textsubscript{2} and its spintronic THz response, demonstrated here for the first time, establishes thickness as a direct and accessible handle on spin-to-charge conversion efficiency, and points toward a broader design strategy for THz emitters based on dimensionally tunable topological materials.

% -------------------------------------------------------
% RESULTS AND DISCUSSION
% -------------------------------------------------------
\section*{Results and Discussion}

\subsection*{Structural and morphological characterization}

\begin{figure}[H]
    \centering
    \includegraphics[width=\linewidth]{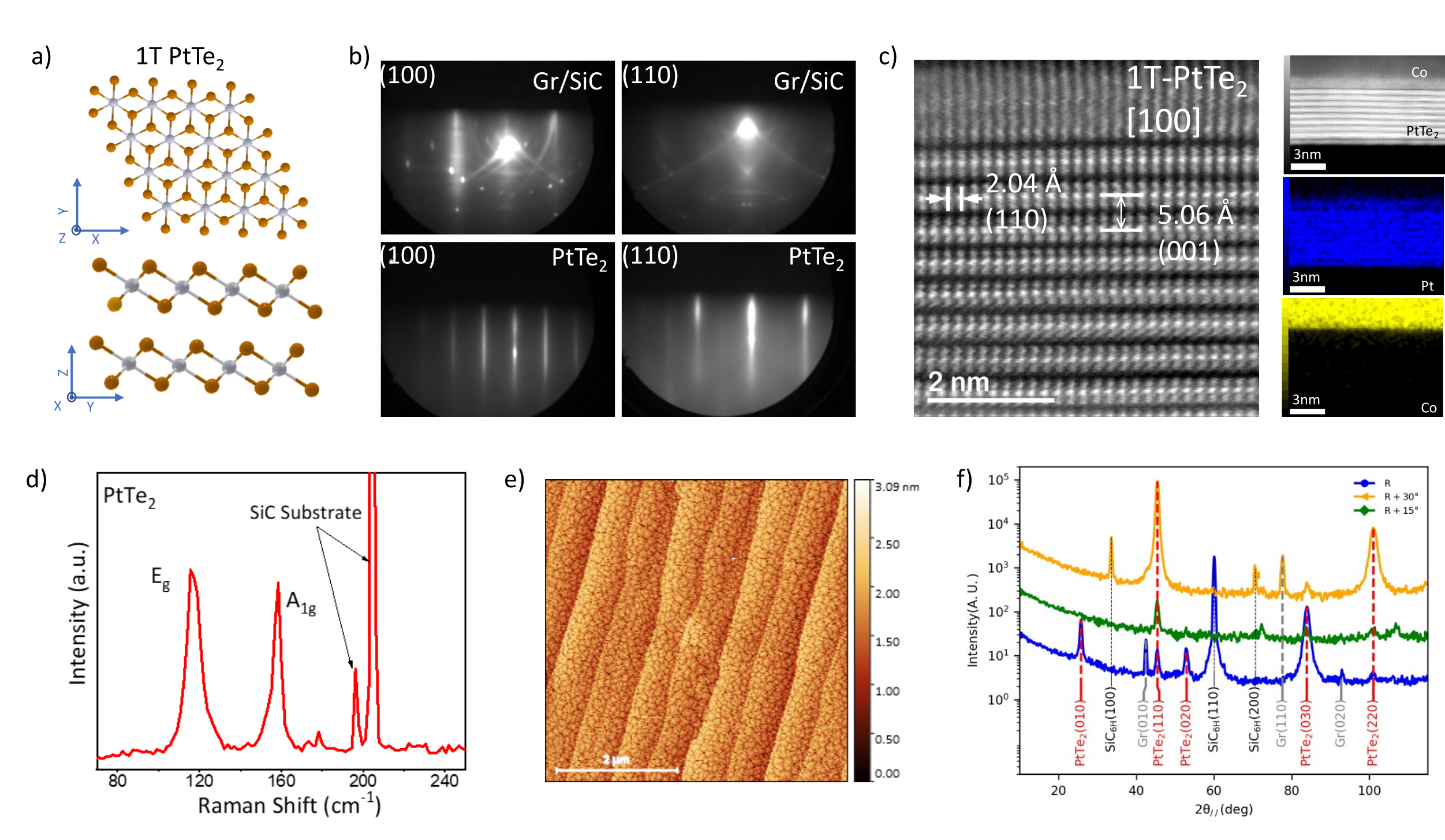}
    \caption{\textbf{Structural and morphological characterization of MBE-grown 1T-PtTe\textsubscript{2} thin films.} (a) Top and side views of the crystal structure of 1T-PtTe\textsubscript{2}. Grey (orange) atoms : Pt (Te) atoms (b) Reflection high-energy electron diffraction (RHEED) patterns recorded along the (100) and (110) azimuths of the Gr/SiC substrate (top) and of the PtTe\textsubscript{2} monolayer grown on top of the substrate (bottom), confirming epitaxial growth and long-range crystalline order.    (c) Left: High-angle annular dark-field scanning transmission electron microscopy (HAADF-STEM) image of 1T-PtTe\textsubscript{2} viewed along the [100] zone axis, revealing the layered atomic arrangement. Scale bar: 2~nm. Right: Cross-sectional energy-dispersive X-ray spectroscopy (EDX) elemental maps showing the distribution of Pt (blue) and Te (yellow), confirming sharp interfaces and stoichiometric composition. Scale bars: 3~nm. (d) Raman spectrum of the PtTe\textsubscript{2} film, exhibiting characteristic phonon modes consistent with the 1T phase. (e) Atomic force microscopy (AFM) image showing the surface morphology of the monolayer PtTe\textsubscript{2} film. Scale bar: 2~\textmu m. (f) Grazing incidence in-plane X-ray diffraction radial spectra along three azimuthal directions (R, R+15°, and R+30°) where R (Reference direction) is that of the most intense peak of the substrate, i.e. SiC(110). Observed peaks are indexed (hkl) according to the material: PtTe$_2$ (red), SiC (black) and graphene (gray).}
    \label{fig:growth}
\end{figure}

Figure~\ref{fig:growth}a shows the 1T structure (space group $P\bar{3}m1$) of PtTe\textsubscript{2}, in which Te-Pt-Te triatomic layers stack along the $c$-axis via van der Waals interactions. High-quality 1T-PtTe\textsubscript{2} thin films were grown on graphene/SiC(0001) substrates by molecular beam epitaxy (see methods section for details). The crystalline quality and epitaxial growth of the films were monitored \textit{in~situ} by reflection high-energy electron diffraction (RHEED). The substrate prior to growth shows sharp, streaky diffraction patterns along both the (100) and (110) azimuths, consistent with the well-ordered graphene surface (Fig.~\ref{fig:growth}b, top). Upon PtTe\textsubscript{2} deposition, the RHEED patterns evolve into a distinct set of streaks corresponding to the PtTe\textsubscript{2} lattice (Fig.~\ref{fig:growth}b, bottom), confirming epitaxial growth and long-range crystalline order across the film. The streak sharpness and absence of transmission spots indicate a smooth, two-dimensional growth mode throughout.

The atomic structure of the films was examined by high-angle annular dark-field scanning transmission electron microscopy (HAADF-STEM). The cross-sectional image along the [100] zone axis (Fig.~\ref{fig:growth}c, left) reveals the characteristic layered arrangement of
alternating Pt and Te atomic planes, with measured interplanar spacings of 2.04~\AA{} along the [110] direction and 5.06~\AA{} along the [001] direction, in close agreement with the bulk lattice parameters of 1T-PtTe\textsubscript{2} ($a = 4.02$~\AA{}, $c = 5.22$~\AA{}).\cite{Furuseth1965} The contrast in the HAADF image, which scales approximately as $Z^{2}$, clearly distinguishes the heavier Pt columns from the Te planes, and the regularity of the stacking confirms the absence of structural disorder or intercalated phases. Cross-sectional EDX elemental maps (Fig.~\ref{fig:growth}c, right) show well seperated Pt (blue) and Co (yellow) signals with sharp boundaries at both the substrate and Co overlayer interfaces, confirming clean interface formation. Further details are available in supporting information and method section.

Phase identification was further confirmed by Raman spectroscopy. The spectrum of the PtTe\textsubscript{2} film (Fig.~\ref{fig:growth}d) exhibits two prominent features: the E$_{2g}$ mode near $\sim$116~cm$^{-1}$, associated with in-plane Te--Pt--Te vibrations, and the A$_{1g}$ mode near $\sim$158~cm$^{-1}$, corresponding to out-of-plane vibrations.\cite{Li2025} Both positions are consistent with previously reported values for 1T-PtTe\textsubscript{2}, and no additional peaks attributable to secondary phases such as PtTe or elemental Te are observed. The surface morphology of the film, assessed by atomic force microscopy (Fig.~\ref{fig:growth}e), shows a smooth continuous film with terraced features that follow the step-and-terrace structure of the underlying SiC substrate. The root-mean-square roughness remains below 0.5~nm over a 5$\times$5~\textmu m$^{2}$ area, confirming the layer-by-layer growth as indicated by RHEED.

Grazing incidence in-plane X-ray diffraction radial scans were recorded along three azimuths, R, R+15° and R+30°, where R stands for the Reference azimuth based on the most intense substrate peak, i.e. SiC(110). They reveal a clear dependence of the PtTe\textsubscript{2} reflection intensities on the sample azimuthal orientation. For the most intense PtTe\textsubscript{2} peaks, i.e. (110), (030) and (220), intensity is visible on all 3 azimuthal scans but with a very strong anisotropy, allowing confirmation of the preferred epitaxial relationship (ref. RHEED) with the graphene and SiC substrate: PtTe\textsubscript{2}(010)//Gr(010)//SiC(110). For the weak intensity reflections, PtTe\textsubscript{2}(010) and (200), the peaks are visible only on the R azimuthal scan, since in the others they are indistinguishable from background fluctuations. This azimuthal anisotropy indicates that the PtTe\textsubscript{2} grains adopt a preferred in-plane crystallographic orientation relative to the graphene/SiC substrate, consistent with van der Waals epitaxy in which the hexagonal PtTe\textsubscript{2} lattice aligns with the underlying graphene symmetry directions. The measured in-plane lattice parameter of $a_{\mathrm{PtTe_2}} = 3.995$\,\AA\ is in close agreement with the bulk reference value of 4.010\,\AA\ (COD 9009116), with the small residual compressive strain of approximately $-0.37\%$ attributed to the epitaxial constraint imposed by the graphene template.

\subsection*{THz emission properties}

\begin{figure}[H]
    \centering
    \includegraphics[width=\linewidth]{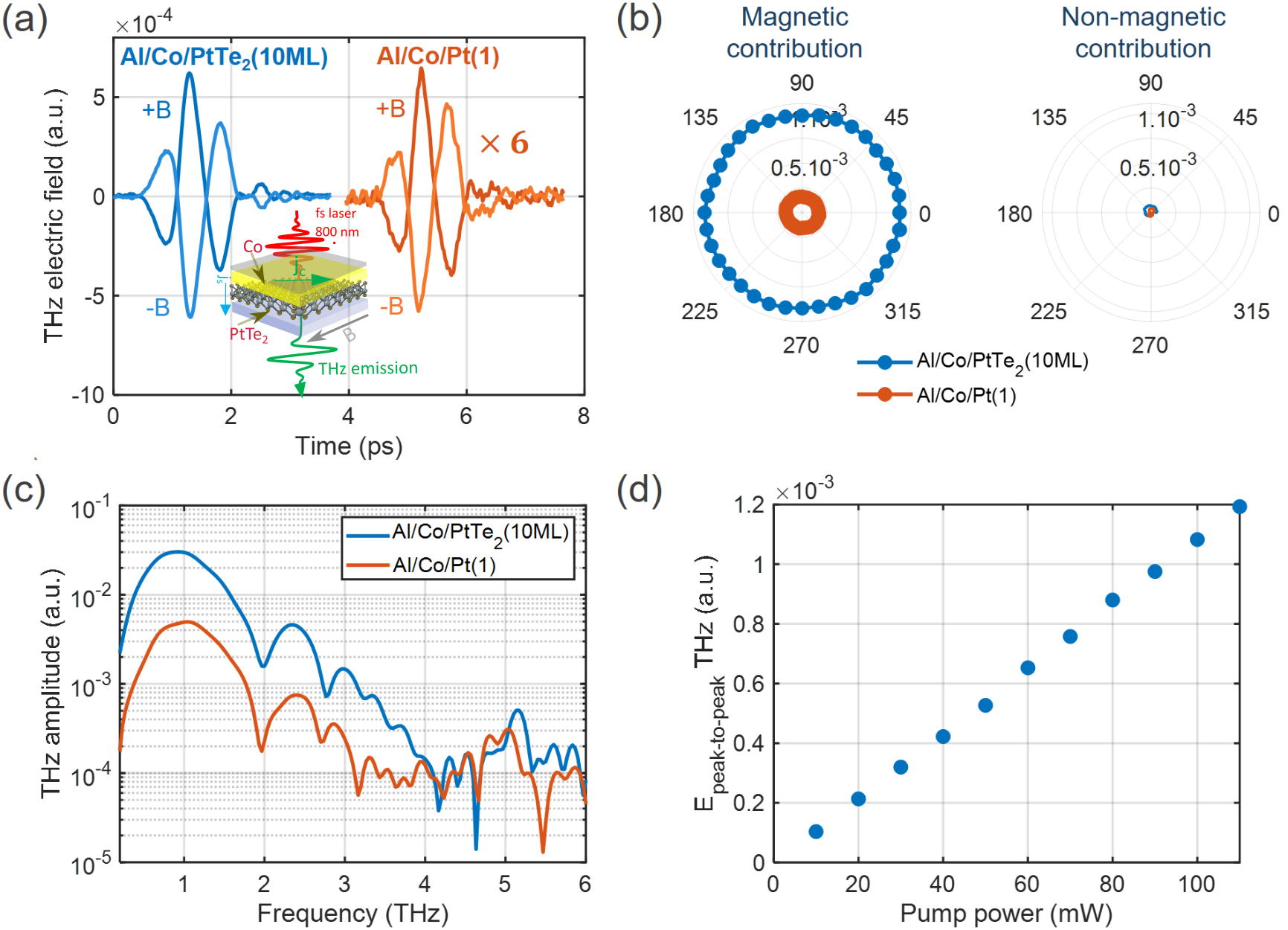}
    \caption{\textbf{PtTe\textsubscript{2} THz emission properties.}
    (a) THz emission time traces for Al(3\,nm)/Co(3\,nm)/PtTe\textsubscript{2}(10\,ML) (blue) compared with the reference sample Al(3\,nm)/Co(3\,nm)/Pt(1\,nm) (red) with equivalent Pt thickness, for small ($\sim$20\,mT) in-plane positive (+$B$) and negative ($-B$) magnetic fields. The 10\,ML PtTe\textsubscript{2} stack has the same conversion sign as Pt and exhibits a sixfold increase in emitted THz electric field compared to the reference. Temporal traces are shifted for clarity. (b) Peak-to-peak amplitude of the magnetic ($E_{+B}-E_{-B})/2$ and non-magnetic ($E_{+B}+E_{-B})/2$ contributions extracted from (a), reported on a polar plot as a function of the sample azimuthal angle. Both stacks show isotropic magnetic THz emission with negligible non-magnetic contribution.
    (c) THz spectrum of the magnetic contribution for the two samples, showing broadband emission with a bandwidth of $\sim$3\,THz, limited by the probe pulse duration and the electro-optic detector. (d) Power dependence of the magnetic peak-to-peak THz emission from Al(3\,nm)/Co(3\,nm)/PtTe\textsubscript{2}(10\,ML), showing linear behavior up to 110\,mW incident optical power.}
    \label{fig:THz}
\end{figure}

After confirming the structural quality of the MBE-grown PtTe\textsubscript{2} films, we turn to their spintronic THz emission properties. Figure~\ref{fig:THz}a shows time-domain THz traces recorded from Al(3\,nm)/Co(3\,nm)/PtTe\textsubscript{2}(10\,ML) and a reference Al(3\,nm)/Co(3\,nm)/Pt(1\,nm) sample under in-plane magnetic fields $+B$ and $-B$. Reversing the applied field inverts the
polarity of the emitted THz waveform in both samples which is the hallmark signature of spin-current-driven emission. The 10\,ML PtTe$_2$ stack generates a THz electric field approximately six times larger than the 1.09 \,nm Pt reference of equivalent Pt content (considering that 10ML of PtTe$_2$ is
equivalent in Pt content to a 1.09 nm-thick pristine Pt layer, based on interlayer spacing, see Fig. 1c), despite the conversion layer being a TMD rather than an elemental heavy metal. The sign of the emission is the same for both samples showing that SCC mechanisms in PtTe\textsubscript{2} and Pt share the same polarity.

To verify that the THz emission is genuinely spin-current-driven and free of other contributions— such as optical rectification, photogalvanic effects, or substrate-related nonlinearities \cite{Li2025,Chen2025} — we deconvolute the THz signal into its magnetic and non-magnetic components by recording traces at each azimuthal angle $\phi$ and extracting $(E_{+B} - E_{-B})/2$ and $(E_{+B} + E_{-B})/2$ respectively. The polar plots in Fig.~\ref{fig:THz}b show that the magnetic contribution is isotropic and dominates the signal for both samples across the full 360° rotation, while the non-magnetic component remains negligible at all angles. Any significant contribution from nonlinear optical effects or photogalvanic currents would produce a characteristic angular dependence, which is not observed here.\cite{Chen2025} We further demonstrate that these properties are also preserved while growing PtTe$_2$ films on other substrates like sapphire (see Figure S1 and S2 of the Supp. Info.).

The frequency-domain spectra of the magnetic THz contribution, shown in Fig.~\ref{fig:THz}c, confirm broadband emission extending to approximately 3\,THz for both stacks, with the bandwidth limited by the probe pulse duration and the phase-matching bandwidth of the electro-optic ZnTe detector rather than by any intrinsic material cutoff. Across the entire spectral range, the PtTe\textsubscript{2} emitter maintains its advantage over the Pt reference by a rough constant factor of 6. Figure~\ref{fig:THz}d shows the dependence of the peak-to-peak THz amplitude on incident optical pump power for the Al/Co/PtTe\textsubscript{2}(10\,ML) stack. The emission scales linearly with pump power up to the maximum measured value of 110\,mW, with no sign of saturation. This linear behavior confirms that the THz generation process operates in the perturbative regime under our measurement conditions, where the emitted field is proportional to the laser-induced spin current and hence to the number of photoexcited carriers in the Co layer. The absence of nonlinear rolloff also indicates that the PtTe\textsubscript{2} layer does not introduce significant optical absorption or pump-induced heating effects that would otherwise degrade the spin injection efficiency at higher fluences.

\subsection*{Thickness dependence}

\begin{figure}[H]
    \centering
    \includegraphics[width=\linewidth]{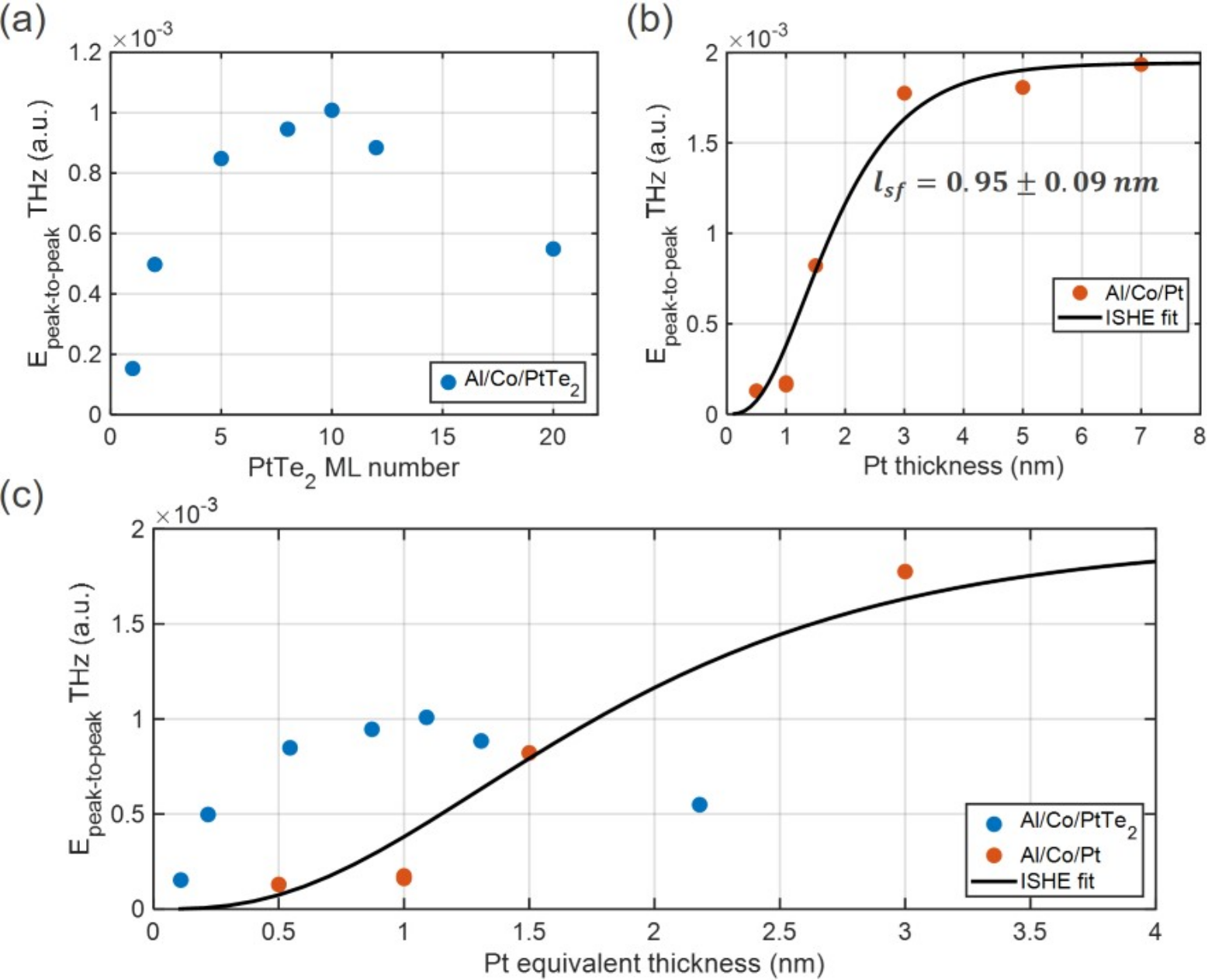}
    \caption{\textbf{Thickness dependence of THz emission.}
    (a) Peak-to-peak amplitude of THz emission from Al/Co/PtTe\textsubscript{2} as a function of PtTe\textsubscript{2} thickness expressed in monolayer number. (b) Peak-to-peak amplitude of THz emission from Al/Co/Pt as a function of Pt thickness (red dots). The black solid line corresponds to the ISHE fit using equation~(1), yielding a spin relaxation length of 0.95\,nm. (c) Comparison between THz emission from Al/Co/PtTe\textsubscript{2} (blue dots) and Al/Co/Pt (red dots) with its ISHE fit, as a function of Pt-equivalent thickness in the spin-to-charge conversion material (PtTe\textsubscript{2} or Pt), considering that 10\,ML of PtTe\textsubscript{2} is equivalent in Pt content to a 1.09 \,nm-thick pristine Pt layer.}
    \label{fig:thickness}
\end{figure}

To understand how the spin-to-charge conversion efficiency of PtTe\textsubscript{2} evolves with its thickness, we measured the peak-to-peak THz amplitude as a function of PtTe\textsubscript{2} layer thickness expressed in monolayer number (Fig.~\ref{fig:thickness}a). The emission is weak but nonzero at 1\,ML, rises steeply through the 2-5\,ML range, reaches a broad maximum near 10\,ML,
and then decreases at 20\,ML. This non-monotonic behavior is qualitatively distinct from what one expects for a simple bulk ISHE material, where emission saturates smoothly on the scale of the spin relaxation length and does not turn over. The initial rise directly tracks the electronic phase evolution established by ARPES from the semiconducting single-layer limit through the thin-film semimetal regime.\cite{Politano2018} The further sharp increase to 10\,ML may be related to the evolution of the material into the fully developed type-II Dirac semimetal phase. Finally, the decrease of THz emission at larger thicknesses reflects growing THz absorption by the increasingly metallic and thicker PtTe\textsubscript{2} layer, which attenuates the emitted pulse before it exits the stack.

As a quantitative benchmark, we measured the Pt thickness dependence on THz emission of Al(3\,nm)/Co(3\,nm)/Pt($x$) reference heterostructures on the same graphene/SiC substrate (Fig.~\ref{fig:thickness}b). The data follow the standard ISHE thickness dependence cleanly, and accordingly fits with the expression\cite{Krishnia2024}:
\begin{equation}
E_{\mathrm{THz}} \propto
\left( \theta_{\mathrm{SHE}} l_{sf} \right)
\frac{
G_s (\rho_{Pt} l_{sf}) \tanh^2 \left( \frac{t_{Pt}}{2 l_{sf}} \right)
}{
1 + G_s (\rho_{Pt} l_{sf}) \coth \left( \frac{t_{Pt}}{l_{sf}} \right)
}
\end{equation}
where $l_{\mathrm{sf}}$ is the spin relaxation length, $G_s$ is a unitless parameter  (set to 2 in the fit) depending on $l_{\mathrm{sf}}$ and the platinum resistivity $\rho_{\mathrm{Pt}}$, and $t_{\mathrm{Pt}}$ is the Pt thickness. This yields a spin relaxation length of $l_{\mathrm{sf}} = 0.95 \pm 0.09$\,nm for Pt, in good agreement with previously reported spin diffusion lengths in sputtered Pt, confirming the reliability of our measurement geometry and normalization procedure.\cite{Tao2018}

The main comparison is shown in Fig.~\ref{fig:thickness}c, where the PtTe\textsubscript{2} and reference data are replotted against a Pt-equivalent thickness axis. This conversion is based on the stoichiometry of the 1T structure: 10\,ML of PtTe\textsubscript{2} contains the same amount of Pt atoms as a 1.09\,nm thick layer of elemental Pt, providing a physically meaningful basis for comparing the two materials on equal footing in terms of Pt content. The Pt data and their ISHE fit (black curve) rise steeply from zero and saturate near 3-4\,nm equivalent thickness, as expected for a material in which SCC is entirely bulk in origin and limited by spin relaxation. The PtTe\textsubscript{2} data, by contrast, already exceed the ISHE prediction at equivalent thicknesses below $\sim$1\,nm of Pt, and sustain emission well above the Pt curve across the full thickness range where the type-II Dirac phase is established. At the optimal thickness of
10\,ML ($\sim$1\,nm Pt-equivalent), the PtTe\textsubscript{2} spin-conversion layers delivers approximately six times the THz amplitude of the same Pt content in elemental form.

This excess emission above the ISHE prediction cannot be accounted for by the bulk inverse spin Hall effect nor by the bulk inverse orbital Hall effect which is even weaker (see Supp. Info. Fig. S6) but points directly to the additional SCC channels unique to the PtTe\textsubscript{2} electronic structure — namely the inverse Rashba-Edelstein contribution from the topological surface states and the local Rashba channel associated with the gapped Dirac cone between the $\Gamma$ and M points. Both of these surface- and interface-related mechanisms contribute most effectively at intermediate thicknesses where the type-II Dirac band structure is established but before bulk THz absorption becomes the dominant limiting factor, consistent with the observed peak near 10\,ML. The subsequent decline toward 20\,ML, which the ISHE model does not predict, further supports this interpretation: as the film thickens, the relative weight of the surface state contribution diminishes and THz reabsorption in the metallic bulk increasingly suppresses the emitted signal.

\subsection*{Electronic Structure and Microscopic Origin of Enhanced Spin-to-Charge Conversion}

\begin{figure}[H]
    \centering
    \includegraphics[width=\textwidth]{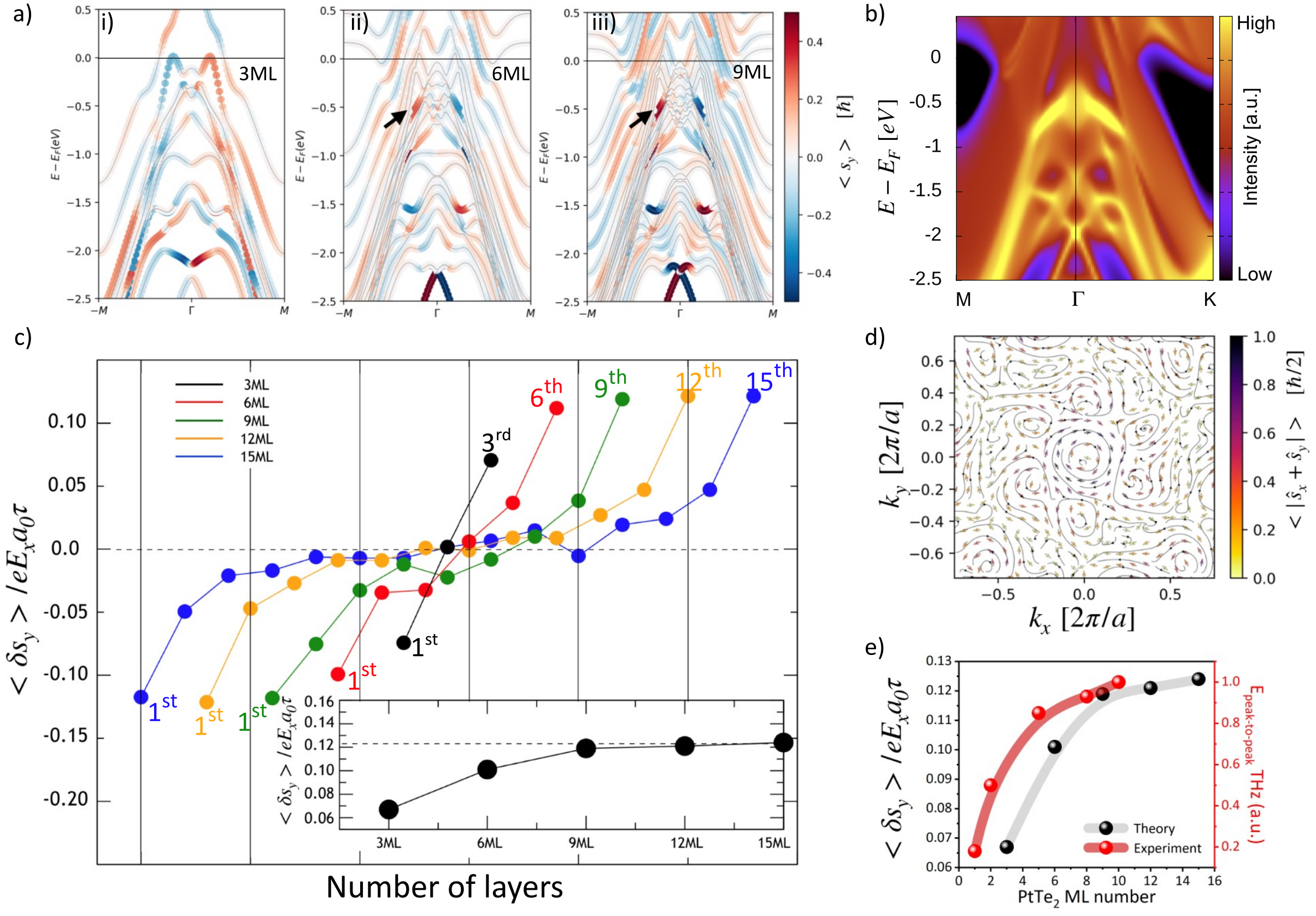}
    \caption{\textbf{Calculated electronic structure and spin-to-charge conversion in PtTe\textsubscript{2}} (a) DFT-calculated spin-resolved band structures of PtTe\textsubscript{2} slabs at representative thicknesses of 3\,ML (i), 6\,ML (ii), and 9\,ML (iii), plotted along the $\overline{\text{M}}$--${\Gamma}$--${\text{M}}$  high-symmetry direction. The color scale indicates the expectation value of the  spin operator $\langle s_y \rangle$ in units of $\hbar$, revealing the progressive development of spin-split bands and topological surface states (marked by black arrows) with increasing thickness. (b) Calculated bulk band structure of PtTe\textsubscript{2} showing the anisotropic spectral weight distribution characteristic of the type-II Dirac semimetal phase, with the white dashed line marking the Fermi level $E_F$. (c) Thickness-dependent spin accumulation $\langle \delta s_y \rangle / eE_x a_0 \tau$ calculated for PtTe\textsubscript{2} slabs of 3, 6, 9, 12, and 15\,ML as a function of layer index across the slab, illustrating the evolution from a weakly polarized to a surface-state-dominated spin response with increasing thickness. Inset: Interfacial spin accumulation as a function of PtTe\textsubscript{2} thickness, saturating near 12\,ML. (d) Spin texture map of the accumulated spin polarization in the $k_x$--$k_y$ plane for 6 ML PtTe\textsubscript{2}. (e) Direct comparison of the thickness-dependent spin accumulation from theory (black dots, left axis) and peak-to-peak THz emission amplitude from experiment (red dots, right axis) as a function of PtTe\textsubscript{2} monolayer number, demonstrating quantitative agreement across the full studied thickness range.}
    \label{fig:theory}
\end{figure}

The experimental THz emission trends described above are underpinned by a systematic evolution of the PtTe\textsubscript{2} electronic structure with thickness, which we then examine through first-principles calculations. Figure~\ref{fig:theory}a shows the spin-resolved band structures computed for PtTe\textsubscript{2} slabs of 3, 6, and 9\,ML along the $\overline{\text{M}}$--${\Gamma}$--${\text{M}}$ direction, with the color scale encoding the transverse spin component $\langle s_y \rangle$. At 3\,ML, the band structure near the Fermi level already exhibits semimetallic character with bands crossing $E_F$, but 
the spin polarization remains weak and the surface states are poorly defined. At this thickness, the top and bottom surface wavefunctions overlap substantially, and the resulting hybridization partially gaps and mixes the surface states, degrading spin--momentum locking and reducing the net spin polarization available for conversion. As the thickness increases to 6 and 9\,ML, this hybridization is progressively suppressed: the surface states sharpen into well-defined dispersive bands with strong spin polarization (marked by black arrows in panels ii and iii of Fig.~\ref{fig:theory}a), and their spin texture becomes increasingly helical and robust. The bulk limit, 
shown in Fig.~\ref{fig:theory}b, represents the endpoint of this evolution~--- a fully developed type-II Dirac semimetal, a strongly tilted Dirac cone, and well-separated topological surface states carrying near-maximal spin--momentum locking. The thickness-dependent band structures of Fig.~\ref{fig:theory}a are therefore best understood as a gradual approach toward this bulk limit, with each added layer contributing to the decoupling and sharpening of the surface states.

Alongside the surface state evolution, the structural inversion symmetry breaking at the ferromagnet/PtTe\textsubscript{2} interface plays an important and thickness-dependent role. As the topological surface states become more localized near the PtTe2/Co interface, their penetration into the bulk is reduced, suppressing finite-size overlap and hybridization effects between opposite surfaces. This improved localization preserves the interfacial spin texture more effectively, leading to a stronger momentum-dependent spin polarization and, consequently, to a more efficient spin-to-charge conversion via the inverse Rashba–Edelstein effect.

The combined effect of these two processes~--- the decoupling and sharpening of topological surface states and the strengthening of interfacial Rashba splitting~--- is quantified through the calculated spin accumulation $\langle \delta s_y \rangle / eE_x a_0 \tau$ shown as a function of layer index across slabs of different thickness in Fig.~\ref{fig:theory}c. At all thicknesses, the spin accumulation is largest at the interface layer (top and bottom) and decays into the bulk of the slab, consistent with the interfacial and surface-state origin of the dominant spin-to-charge conversion channels. The magnitude of this interfacial peak grows monotonically from 3\,ML to 10\,ML before beginning to saturate near 12\,ML (inset of Fig.~\ref{fig:theory}c), directly mirroring the experimental THz emission trend. The spin texture map of Fig.~\ref{fig:theory}d shows the momentum-space distribution of the accumulated spin polarization for a 6 ML PtTe\textsubscript{2} slab thickness. The complex winding pattern, with vortex-like features distributed across the Brillouin zone, reflects the superposition of contributions from the helical topological surface states and the Rashba-splitted interfacial bands~- both activated by the same broken inversion symmetry at the Co/PtTe\textsubscript{2} interface but carrying distinct momentum-space signatures.

The quantitative connection between theory and experiment is made explicit in Fig.~\ref{fig:theory}e, which overlays the calculated interfacial spin accumulation (black dots, left axis) and the measured peak-to-peak THz amplitude (red dots, right axis) as a function of PtTe\textsubscript{2} monolayer number. Both quantities follow the same sub-linear growth with thickness, rising steeply in the 1--6\,ML range and saturating gradually beyond 10\,ML. This quantitative agreement across the full thickness range provides strong evidence that the thickness-dependent THz emission enhancement is directly governed by the growth of interfacial spin accumulation, driven by the progressive development of topological surface states and Rashba-splitted interfacial bands as the type-II Dirac semimetal phase is established. It further confirms that the dominant spin-to-charge conversion mechanism in PtTe\textsubscript{2}-based spintronic emitters is not simply a scaled version of the bulk inverse spin Hall effect, but reflects the richer multi-channel physics that becomes fully accessible only once the film thickness is sufficient to support a well-developed type-II Dirac band structure.
% -------------------------------------------------------
% CONCLUSION  (to be filled)
% -------------------------------------------------------
\section*{Conclusion}

In summary, we have demonstrated that the spintronic Terahertz emission from MBE-grown PtTe\textsubscript{2} thin films is not a fixed material property but an electronically tunable one, controlled directly by film thickness. By growing PtTe\textsubscript{2} with single-monolayer precision across the range from 1 to 20\,ML, we have mapped the THz emission response onto the electronic phase diagram of this material from a semiconducting single layer that emits very limited THz signal, through a thin-film semimetal regime where emission turns on sharply, to a fully developed type-II Dirac semimetal phase where the combination of topological surface states and interfacial Rashba splitting drives the emission to a level approximately six times that of an equivalent Pt thickness. First-principles calculations of the thickness-dependent spin accumulation reproduce the experimental trend quantitatively, identifying the progressive decoupling of surface states and the strengthening of interfacial spin--orbit coupling as the two microscopic mechanisms responsible for the enhancement. Taken together, these results establish thickness as a direct and physically transparent handle on spin-to-charge conversion efficiency in PtTe\textsubscript{2}, and demonstrate that the rich phase diagram of dimensionally tunable van der Waals semimetals, which was long studied in the context of surface physics, can be directly harnessed for the design of next-generation spintronic Terahertz emitters or spin-orbit torque magnetic memories made of van der Waals materials.

% -------------------------------------------------------
% MATERIALS AND METHODS  (to be filled)
% -------------------------------------------------------
\section*{Materials and Methods}

\noindent \textbf{MBE Growth.}
PtTe$_{2}$ thin films were grown epitaxially on two substrate platforms: single-layer graphene/SiC and c-plane sapphire (0001). The graphene/SiC(0001) substrate was prepared by CVD in a hydrogen atmosphere ~\cite{Mastropasqua2025} under conditions leading to the formation of a single layer of graphene on a buffer layer. For both substrates, the substrate was thermally degassed at 800\,°C for 30\,minutes prior to growth. For graphene/SiC substrates, the substrate was subsequently cooled to the PtTe$_{2}$ growth temperature of 285\,°C. For sapphire substrates, the growth temperature was set to 450\,°C. In both cases, platinum was deposited at a rate of $\approx$0.1\,\AA/min under a tellurium background pressure of 5\,$\times$\,10$^{-8}$\,mbar. Following deposition, the films grown on graphene/SiC were annealed in situ at 405\,°C for 10\,minutes, while those grown on sapphire were annealed at 600\,°C for 5\,minutes, in both cases to promote crystalline order. The crystalline quality of the PtTe$_{2}$ films on both substrates were confirmed by in situ RHEED measurements; the RHEED patterns for PtTe$_{2}$ grown on sapphire are shown in Supplementary Figure ~S1. After cooling to room temperature, a 3\,nm Co ferromagnetic layer was deposited in the same MBE chamber without breaking vacuum, at a rate of 0.8\,\AA/s, followed immediately by a 3\,nm Al capping layer deposited at 1\,\AA/s to prevent oxidation of the ferromagnetic film.

\noindent \textbf{X-ray Diffraction.}
Structural characterization by grazing incidence X-ray diffraction (GIXRD) was carried out on a SmartLab Rigaku diffractometer fitted with a copper rotating anode source (K$_{\alpha}$ = 1.54\,\AA) running at 45\,kV and 200\,mA. In-plane parallel collimators with an angular resolution of 0.5° were mounted on both the incident and diffracted beam paths.

\noindent \textbf{Scanning Transmission Electron Microscopy.}
Cross-sectional TEM lamellae were extracted using a Scios 2 DualBeam (Thermo Fisher Scientific) FIB-SEM instrument with a Ga$^+$ focused ion beam. Prior to milling, the sample surface was protected by a vacuum-evaporated carbon layer followed by a platinum layer deposited by combined electron- and ion-beam-induced deposition within the Scios 2 system. Lamella thinning was carried out in three successive stages at 30\,kV, 5\,kV, and a final low-energy polish at 2\,kV to minimize surface damage and amorphization. Structural, morphological, and compositional analyses were performed on a cold-field-emission spherical-aberration-corrected analytical transmission electron microscope JEM-ARM 200cF (JEOL, Japan) operated at 200\,kV. Energy-dispersive X-ray spectroscopy was conducted using a 100\,mm$^{2}$ silicon drift detector JED-2300 (JEOL, Japan) integrated with the same instrument.

\noindent \textbf{Terahertz Emission Measurements.}
THz emission measurements were performed in transmission geometry with optical excitation from the substrate side at normal incidence. Linearly polarized pump pulses of 80\,fs duration centered at 800\,nm were delivered by a Yb-doped solid-state femtosecond laser system coupled to an optical parametric amplifier, operating at a repetition rate of 150\,kHz with a beam waist of 250\,$\mu$m at the sample position. A small in-plane magnetic field of approximately 20\,mT was applied during measurements to saturate the Co layer magnetization. The emitted THz pulses were collected and focused onto a 200\,$\mu$m-thick $\langle$110$\rangle$ ZnTe electro-optic crystal oriented orthogonally to the applied magnetic field direction, and detected by free-space electro-optic sampling~\cite{} using the 130\,fs probe pulses at 1030\,nm from the main oscillator output.

\noindent \textbf{DFT calculations:}
Please find the details in DFT section of Supplementary information.

% -------------------------------------------------------
% ACKNOWLEDGEMENTS
% -------------------------------------------------------
\section*{Acknowledgements}
The authors acknowledge the support from the European Union’s Horizon 2020 research and innovation Programme under grant agreements No 881603 (Graphene Flagship) and No 101099552 (EIC Pathfinder PLASNANO). The French National Research Agency (ANR) is acknowledged for its support through the ANR-18-CE24-0007 MAGICVALLEY, ANR-22-CE09-0012-03 COME ON, and ESR/EQUIPEX+ ANR-21-ESRE-0025 2D-MAG projects. The LANEF framework (No. ANR-10-LABX-0051) is acknowledged for its support with mutualized infrastructure. The authors also acknowledge the France 2030 government investment plan PEPR Electronics through the ANR-22-PEEL-0011 ADICT project and PEPR SPIN through the ANR-22-EXSP-0003 TOAST, ANR-22-EXSP-0007 SPINMAT and ANR-22-EXSP-000915 SPINTHEORY projects. Dr. Eva Desgu\'e from Thales Research \& Technology (Palaiseau, France) is acknowledged for fruitful discussions.

\vspace{1 cm}

\noindent The authors declare no competing interests.

\bibliography{references}
\pagebreak
\end{document}